\documentclass[fleqn,usenatbib]{mnras}

\usepackage[T1]{fontenc}
\DeclareRobustCommand{\VAN}[3]{#2}
\let\VANthebibliography\thebibliography
\def\thebibliography{\DeclareRobustCommand{\VAN}[3]{##3}\VANthebibliography}
\usepackage{soul,color}

\usepackage{graphicx}	
\usepackage{amsmath}	
\usepackage{amssymb}	
\usepackage{newtxtext,newtxmath}

\newcommand{\Msun}{\, \mathrm{M}_{\odot}}

\title[Baryon-driven decontraction in MW mass haloes]{Baryon-driven decontraction in Milky Way mass haloes}

\author[Victor J. Forouhar Moreno et al.]{
Victor J. Forouhar Moreno,$^{1}$\thanks{E-mail: victor.j.forouhar@durham.ac.uk}
Alejandro Ben\'itez-Llambay ,$^{2}$
Shaun Cole$^{1}$
and Carlos Frenk$^{1}$
\\
$^{1}$Institute for Computational Cosmology, Department of Physics, Durham University, Durham DH1 3LE, UK\\
$^{2}$University of Milano-Bicocca, Piazza della Scienza, 3, 20126 Milano MI, Italy
}

\date{Accepted XXX. Received YYY; in original form ZZZ}

\pubyear{2021}

\begin{document}
\label{firstpage}
\pagerange{\pageref{firstpage}--\pageref{lastpage}}
\maketitle

\begin{abstract}
  We select a sample of Milky Way (MW) mass haloes from a high-resolution
  version of the EAGLE simulation to study their inner dark matter (DM) content and how baryons alter it.
  As in previous studies, we find that all haloes are more massive at the centre
  compared to their DM-only (DMO) counterparts at the present day as
  a result of the dissipational collapse of baryons during the
  assembly of the galaxy. However, we identify two processes that can
  reduce the central halo mass during the evolution of the
  galaxy. Firstly, gas blowouts induced by AGN feedback can lead to a substantial
  decrease of the central DM mass. Secondly, the formation of
  a stellar bar and its interaction with the DM can induce a secular
  expansion of the halo; the rate at which DM is evacuated
  from the central region by this process is related to the average
  bar strength and the timescale on which it acts determines how
  much the halo has decontracted. Although the inner regions of the haloes we have
  investigated are still more massive than their DMO counterparts at
  $z = 0$, they are significantly less massive than in the
  past and less massive than expected from the classic adiabatic contraction
  model. Since the MW has both a central supermassive black hole and a bar, the extent to which its halo has contracted 
  is uncertain. This may affect estimates of the mass of the MW halo and of the expected signals in direct and indirect DM detection experiments.
\end{abstract}

\begin{keywords}
dark matter -- galaxies: evolution -- galaxies: bar
\end{keywords}



\section{Introduction}

Structure formation in a cold dark matter (CDM) universe proceeds in a
hierarchical bottom-up manner. Small-scale overdensities are the first
to decouple from the Hubble flow and undergo gravitational
collapse. Their subsequent growth is driven by mergers with
neighbouring structures and diffuse, smooth mass accretion from the
surroundings. The end result is a bound, virialised halo of dark
matter (DM) that can host a galaxy at its centre if it is massive
enough \citep{White.1978, Benitez-Llambay.2020}. N-body simulations that model the
dark matter and baryons as a single collisionless fluid predict DM density profiles with shapes that are roughly independent of halo
mass, cosmological parameters and the primordial fluctuation power
spectrum \citep{Navarro.1996b, Wang.2020}. These density profiles are
well fitted by the two parameter Navarro-Frenk-White
\citep{Navarro.1996b,Navarro.1997} profile:
\begin{equation}
    \rho_{\rm NFW}(r) = \dfrac{\rho_{0}}{\dfrac{r}{r_{s}}\Big(1+ \dfrac{r}{r_{s}} \Big)^{2}} \, ,
\end{equation}
although more recent, higher resolution simulations suggest that the
three parameter \cite{Einasto.1965} profile provides an even better
fit \citep{Navarro.2004}. The two parameters of the NFW profile are
related to the halo virial mass and its concentration, both of which
are tightly correlated. This is a consequence of the mass dependence
of the formation epoch of haloes, as a result of which less massive
halos typically have greater concentrations than more massive ones,
reflecting the fact that they undergo gravitational collapse earlier,
when the universe was correspondingly denser.

The ubiquitous prediction of centrally divergent halo density profiles
in CDM offers a test of whether dark matter is cold or not. This has
motivated numerous studies comparing observations of the inferred DM
density profiles to theoretical predictions on a wide range of scales, 
from dwarf galaxies
\citep[e.g][]{Burkert.1995,Agnello.2012,Oh.2015,Walker.2011,Strigari.2010}
to rich galaxy clusters
\citep[e.g.][]{Sand.2002,Umetsu.2017,He.2020}. The comparisons are largely based on predictions stemming
from purely collisionless dark matter-only (DMO) N-body simulations,
where non-linear effects produced by baryons are unaccounted
for. Processes associated with the formation and evolution of galaxies
can have measurable effects on the structure of the DM haloes hosting
them, such as changing the distribution of dark matter or
redistributing angular momentum
(e.g. \citealt{Zavala.2008,Schaller.2015,Chan.2015}). The increasing
availability of hydrodynamical cosmological simulations able both to
reproduce many measured galaxy population statistics and have
sufficient resolution to probe the galaxy-scale distribution of DM
enables a more meaningful comparison between theory and observations
(for a review, see \citealt{Somerville.2015}). These simulations use
subgrid prescriptions to model processes such as star formation, gas
cooling, feedback due to supernovae and AGN. The interplay between
these processes leads to a complex and rich phenomenology that is
missing in DMO simulations.

Initially, gas near a growing DM halo is dragged in due to the
deepening gravitational potential well, shock heated and, if it can cool
efficiently, it will sink towards the centre where star formation can
commence once the gas density is large enough \citep{White.1978}. The dissipative
collapse of gas and its assembly in the central regions of the
halo deepens the potential well, inducing a contraction of the DM halo
and enhancing the central density compared to the DMO counterpart. The
effectiveness of this response depends on a number of properties, such
as the mass of the central galaxy, the assembly history of the halo
and the phase-space distribution of DM particles
\citep{Abadi.2010,Dutton.2016,Artale.2019}. The first analytical
models used to estimate this response assumed `adiabatic
contraction' and circular orbits \citep{Blumenthal.1986,Ryden.1987}
but were later expanded to take into account the orbital
eccentricities of typical dark matter particles
\citep{Gnedin.2004}. More recently, there have been a number of
extensions based on empirical fits to the measured response in
hydrodynamical N-body simulations \citep{Cautun.2020}, as well as
orbital phase space modelling using integrals of motion
\citep{Callingham.2020}.

The assembly of gas and stars has other effects that are in direct competition with the contraction of the halo. The sudden expulsion, driven by supernovae explosions, of gas that had previously accumulated slowly at the centre of a dwarf galaxy halo  can cause the central regions of the halo to expand \citep{Navarro.1996a}. This can occur in a single disruptive event or in a series of more moderate perturbations that drive oscillations in the gravitational potential \citep{Read.2005, Pontzen.2014}. On cluster scales a similar outcome can result from powerful AGN-driven outbursts \citep{Martizzi.2013}.
Similarly, dynamical friction exerted on infalling gas clumps by dwarf-scale haloes \citep{El-Zant.2001,Mashchenko.2006} or merging galaxies in cluster-scale haloes \citep{El-Zant.2004,Laporte.2012} could also lower the central dark matter density.

The details of how star formation is modelled determines the degree of gravitational coupling between the gas and the DM halo, and thus influences how effectively gas blowouts can alter the inner contents of dark matter haloes \citep{Benitez-Llambay.2019}. This explains differences in the predicted density profiles of dwarf
galaxies between simulations employing high density thresholds and those employing lower ones. Whilst the former are able to accumulate sufficient quantities of gas in the central regions of the halo prior
to the gas being blown out, low density thresholds never reach this
point. Another important aspect of gravitational perturbations is the
timescale on which they operate. As discussed in the Appendix of
\citet{Benitez-Llambay.2019}, perturbations that last longer, compared to the typical dynamical time of the shell, remove DM more effectively.  If the
perturbation timescale is sufficiently long, the effectiveness of each individual perturbation in heating the DM becomes maximal. This means that the
integrated effect is solely dependent on the number of such perturbations

Finally, torques exerted by non-axisymmetric features are able to
redistribute angular momentum between baryons and DM, as well as
within the galaxy itself
\citep{Lynden-Bell.1972,Lynden-Bell.1979}. One example are stellar
bars, which are present in a significant fraction of nearby spiral galaxies
\citep{Eskridge.2000,Sheth.2008,Skibba.2012,Buta.2015} and in our own
Milky Way (MW) \citep{Binney.1991,Weiland.1994, Dwek.1995, Ness.2016}. The
formation of a bar can be driven by an instability resulting from a
kinematically cold and gravitationally important disc, as appreciated
in early N-body simulations (e.g \citealt{Miller.1970,Hohl.1971}). An
alternative bar formation mechanism relies on external triggers such
as tidal interactions caused by a close flyby or a merger
\citep{Noguchi.1987, Lokas.2016, Martinez-Valpuesta.2017}. These
processes can also reconstitute previously existing bars
\citep{Berentzen.2004}.

The subsequent evolution of the bar is determined by exchange of
angular momentum, which can lead to its strengthening and lengthening
\citep{Athanassoula.2003}. This exchange occurs near orbital
resonances. While some authors argue  that as many as $10^{8}$ particles are required to model the resonances (e.g. \citealt{Weinberg.2007a, Weinberg.2007b, Ceverino.2007}), others find less stringent conditions on the grounds that a time-evolving bar pattern speed  broadens the resonant regions (e.g. \citealt{Sellwood.2006}).
The net flow of angular momentum depends on the dynamical and spatial
properties of the constituent components of the system (for a review,
see \citealt{Athanassoula.2013}). The regions of the disc within the
corotation radius of a bar lose it, whereas those beyond gain it. On
the other hand, spheroidal components such as the DM halo and the
stellar bulge are only able to acquire it. Consequently, bar driven
changes in the distribution of angular momentum can change the
structural properties of discs \citep{Debattista.2006}, cause
classical bulges to acquire net rotation
\citep{Saha.2012,Kataria.2019} and alter the central density of DM
haloes \citep{Weinberg.2002,Holley-Bockelmann.2005,Sellwood.2008,Dubinski.2009,Algorry.2017}

The efficiency with which all these different processes are able to alter the
central density of dark matter depends largely on the mass scale under consideration. For example, the small amount of baryons collected at the centre of very faint dwarfs cannot alter significantly the inner DM content of their host halo. On the other hand, if too many baryons end up locked in stars in larger haloes, the DM contracts in response to them. There is thus a narrow range in mass in which supernovae-driven gas blowouts are effective at driving DM mass out \citep[e.g.][]{Di_Cintio.2014,Tollet.2016}. It is then common practice to assume that larger haloes, particularly those with mass comparable to that of our Milky Way, are only subject to the contraction caused by baryons, ignoring altogether the competing effects caused by other processes, such as AGN-driven outflows of gas, or the presence of a massive bar at the centre.

In this paper we revisit these ideas using a high-resolution hydrodynamical simulation of the EAGLE project. In particular, we study in detail the time evolution of the inner DM content of a sample of Milky Way-mass haloes and search for events that alter it. Understanding the role of baryons in these haloes is important for a wide range of applications, from mass estimates of our Milky Way \citep{Cautun.2020} to informing direct and indirect searches for 
dark matter \citep{Calore.2015, Bozorgnia.2016, Bozorgnia.2017,Schaller.2016}.

The paper is structured as follows. Section 2 describes the
simulations as well as our selection of a sample of galaxies for
analysis. Section 3 presents our results, focusing first on gas
blowouts and then on stellar bars, as well as on the resulting
contraction and expansion of the central regions of the halo. Our
conclusions are presented in Section~4.

\section{Simulations}

In this section we give an overview of the EAGLE simulations used in this work and describe the selection of our halo sample.

\subsection{The code}

The EAGLE project \citep{Schaye.2015,Crain.2015} is a suite of
hydrodynamical cosmological simulations that follow the formation and
evolution of cosmic structure from $\Lambda$CDM initial conditions
assuming the cosmological parameter values from
\citet{Planck_Collaboration.2014}. They were performed using a
modified version of the P-Gadget3 code \citep{Volker.2005} that
incorporates subgrid prescriptions for the physics relevant to galaxy
formation and evolution: radiative cooling \citep{Wiersma.2009},
photoheating, star formation and evolution
\citep{Schaye.2004,Schaye.2008}, stellar feedback
\citep{Dalla_Vecchia.2012}, black hole seeding
\citep{Springel.2005,Booth.2009}, its subsequent growth and
stochastic, thermal AGN feedback. The values of the parameters used in
modelling these processes were set by requiring a good match to the
observed $z = 0.1$ galaxy stellar mass function, the distribution of
galaxy sizes and the amplitude of the central black hole mass {\em vs}
stellar mass relation. Once calibrated in this way, EAGLE reproduces a
number of population statistics \citep{Schaller.2015,Ludlow.2017}.

In this work we use the higher mass resolution version of EAGLE (see
\citealt{Crain.2015} for details), in which the subgrid physics
parameters were recalibrated to account for the increased mass
resolution. This simulation follows $2 \times 752^{3}$ particles in a
volume 25~Mpc on a side. This resolution corresponds to dark matter and
gas particle masses of $1.21 \times 10^{6} \, \Msun$ and
$2.26\times10^{5}\,\Msun$, respectively. The maximum physical
Plummer-equivalent gravitational softening length is 325 parsecs. There are a
total of 405 temporal outputs between redshifts $z = 20$
and $z = 0$, corresponding to a time resolution of $\sim$60\,Myrs. This provides adequate time resolution to study the
processes of interest in this work.

To identify cosmic structures, we assign particles into distinct
groups according to the friends-of-friends (FoF) percolation algorithm
\citep{Davis.1985}. Each group is made up of particles that are within
0.2 times the mean interparticle separation from one
another. Gravitationally bound substructure is found with the SUBFIND algorithm
\citep{Springel.2001}, which, using particle velocity and position information, 
identifies self-bound structures within a larger FoF group. We follow
the time evolution of the SUBFIND groups by identifying their main
progenitor. This is achieved by cross-matching a subset of the most
bound particles between consecutive time outputs.

\subsection{Sample Selection}

Since we are interested in the central parts of dark matter haloes
similar to the Milky Way's, we restrict our analysis to haloes of mass
$M_{200}$\footnote{$M_{200}$ is defined as the mass contained within
  a sphere of mean density  200 times the critical density of the universe.} at $z = 0$ in the range $0.5 -2.5\times 10^{12} \,
\Msun$. This encompasses recent observational estimates of the Milky
Way's halo mass \citep{Callingham.2019,Cautun.2020}. A total of 45
haloes satisfying this criterion were identified in the hydrodynamical
simulation. Their stellar
masses are shown in the top panel of
Fig.~\ref{figure_1}. The central galaxies exhibit a wide range of masses; most are more massive than
$10^{10} \, \Msun$, but they are typically less massive than 
the Milky Way. This is because the stellar mass
to halo mass relation in EAGLE falls short in the halo mass range of
interest, compared to abundance matching results 
(e.g. \citealt{Moster.2013}). This is related to an underestimate of
the galaxy stellar mass function around the knee  \citep{Schaye.2015}.

Their DMO counterparts were found using the particle ID
information for a subset of the most bound particles in the
hydrodynamical and DMO simulations. The halo centres were found using
the shrinking-spheres algorithm \citep{Power.2003}, run only on the
dark matter particle
distribution. 

\section{Results}

We begin with an overview of the central DM mass evolution of the
selected halo sample. This is followed by a qualitative assessment at the
evolution of four different haloes that illustrate the variety of baryon
processes that change their innermost DM and stellar
content. Finally, we discuss and characterise each of these
processes, namely, an initial halo contraction due to the accumulation
of baryons at the centre, which can then be followed by an expansion caused by AGN-induced gas blowouts, stellar bars or a combination of both.

\subsection{Overview}

\begin{figure}
    \centering
    \includegraphics{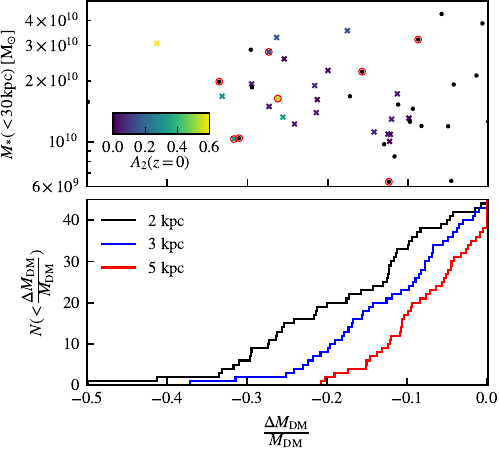}
    \caption{The
      top panel shows the stellar mass of the central galaxies in the selected halo sample, measured within a spherical aperture of 30~kpc, as a function of the fraction
      of DM mass loss within 2~kpc of the halo centre,
      $M_{\rm DM}(z = 0) / M_{\rm DM}(z_{\rm peak}) - 1 $. The crosses
      indicate haloes that exhibit a monotonic decrease in mass with
      time; the dots are the rest of the halo sample.The former are colour-coded according to how prominent their stellar bars are at $z=0$, measured  using the method described in \S3.3. Galaxies that experienced major AGN outbursts at least once during their evolution are highlighted by the red circles. The bottom panel shows the cumulative distribution of.
      the fractional mass loss in the same halo sample, measured
      within spherical apertures of 2~kpc (black), 3~kpc (blue) and
      5~kpc (red). Note that fewer haloes experience a mass loss when
      considering larger apertures, hence why the y-axis intercept of
      the bottom panel changes.}
    \label{figure_1}
\end{figure}
\begin{figure*}
    \centering
    \includegraphics{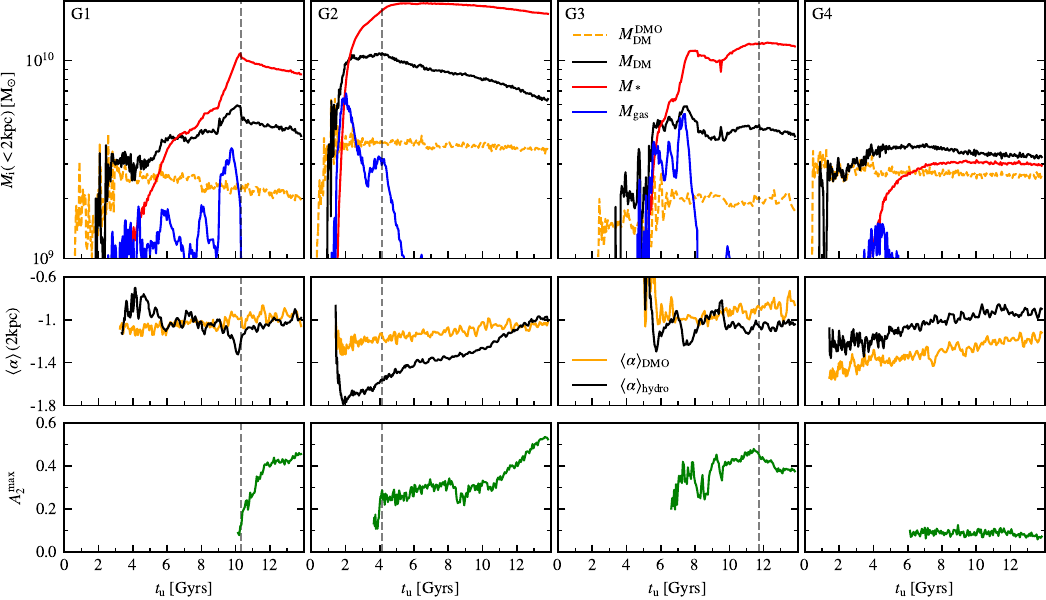}
    \caption{Time evolution of the central dark matter, gas and
      stellar content of four Milky Way mass haloes, together with the
      evolution of their measured bar strengths. Each column is for a
      different halo, with the top panels showing the enclosed DM
      (black), gas (blue) and stellar (red) masses within a 2~kpc
      spherical aperture. For comparison, the DM mass enclosed within
      2~kpc in the DMO counterpart is shown by the dashed orange
      line. The middle panels show the (smoothed) evolution of the DM density profile slope at 2~kpc, both for the DMO (orange) and hydrodynamical (black) counterparts. The bottom panels show the time evolution of the bar strength of these
      galaxies. For haloes G1 and G2, this is only shown
      from times shortly before a visually recognisable bar had 
      formed. Halo G3 had large values of $A^{\rm max}_{2}$ even
      before an established bar formed, which happened shortly before
      the observed peak in enclosed DM mass. Halo G4 never formed a
      bar, hence the low values of $A_{2}$. The vertical dashed lines
      indicate the times used to determine $M_{\rm DM}(z_{\rm peak})$.}
    \label{figure_2}
\end{figure*}
To determine whether there has been a decrease in the central mass of
DM over time, we follow its time evolution within several apertures. The evolution within a 2 kpc aperture is shown for a
few examples in Fig.~\ref{figure_2}.  We locate the time when the dark matter content peaks and compare it to the present-day value. To this end we define the fractional mass loss as
$\Delta M_{\rm DM}/M_{\rm DM} \equiv M_{\rm DM}(z = 0)/M_{\rm
  DM}(z_{\rm peak}) -1$. To prevent fluctuations caused by merger events, which can cause the DM mass to fluctuate for a short period of time, we apply a linear Sagvol-Kolmogorov convolution to smooth
out the evolution. We only consider peaks that are not immediately
followed by a local minimum. This helps prevent transient peaks caused
by mergers with other halos, which would otherwise boost the value of
$\Delta M_{\rm DM} / M_{\rm DM}$.  In practice, this may underestimate
the expansion for a subset of haloes, as illustrated by halo G3 in
Fig.~\ref{figure_2} for which the local maximum at
$t_{\rm u} \sim 11~\mathrm{Gyrs}$ is used instead of the global maximum
at $t_{\rm u} \sim 7.5~\mathrm{Gyrs}$.

The bottom panel of Fig.~\ref{figure_1}
shows the cumulative number of simulated MW-mass haloes that have lost a fractional mass $<f$ within three different spherical apertures: 2~kpc, 3~kpc and
5~kpc. Clearly, the mass loss does not depend on the galaxy stellar mass, as demonstrated in the top panel of the same figure. Although there is a wide range of fractional mass loss values, no halo has lost more than 50\% of its peak dark matter mass, even within 2~kpc from the centre. Nonetheless, half of all haloes have lost more than 16\% of their peak dark matter mass within 2kpc, with only one reaching the peak at $z=0$. The DM mass loss decreases when considering larger apertures: half of the haloes considered here lose more than 11\% and 7\% their peak dark matter mass within 3~kpc and 5~kpc, respectively. Nonetheless, there are still a number of haloes that exhibit
a more significant decrease ($\sim 20\%$) of DM mass even at 5~kpc. To
investigate the reason behind the expansion, we follow the
evolution of the dark matter halo and its associated central baryonic
content.

Four representative examples are shown in
Fig.~\ref{figure_2}, where the time evolution of the
halo dark matter, gas and stellar mass within 2 kpc are shown in the
top panels by different colour lines, as indicated in the
legend. These examples are chosen to illustrate the diverse evolution of the mass content that characterises the haloes in our sample. Some exhibit only a secular decrease of DM
mass over time (G2); others experience an additional, sudden mass loss
event (G1). There are those that remain virtually unaltered throughout their lifetime (G4) and
those which have a more complicated assembly history (G3). Their evolution is
compared to their DMO counterparts, whose inner dark matter content,
$M^{\rm DMO}_{\rm DM} = (1-f_{\rm b})M^{\rm DMO}_{\rm tot}$, is shown
by the black dashed lines.

In all cases, the relative difference between the enclosed dark matter in the hydrodynamical and DMO versions of the same haloes evolves with time. Their values are similar at large redshifts, but start to diverge once gas and stars populate the inner regions of haloes. This is evidence for the contraction of the halo induced by baryons. By $z = 0$, all the haloes in the hydrodynamical simulation are more massive at the centre than their DMO twins. Nonetheless, it is evident from haloes G1, G2 and G3 that their central DM mass in the hydrodynamical simulation evolves non-monotonically, leading to a decontraction at late times. We measure the slope of the DM density profile by fitting a power law, $\rho \propto r^{\alpha}$, to its central distribution. The time evolution of this quantity is shown in the middle panels of Fig. \ref{figure_2}. None of the haloes considered here show signs of a significant flattening at $\sim$ 2~kpc, indicating that neither AGN nor bars make cores on these scales in our simulations Moreover, there is very little difference in the slopes between the hydrodynamical and DMO versions of the same halo, despite large differences in the enclosed mass at 2~kpc. The local density slope is not an adequate metric for quantifying how contracted a halo is.

Although the physical mechanisms driving these mass changes will be discussed in detail in
the following subsections, we present here a  qualitative discussion on the evolution of the haloes shown in Fig. \ref{figure_2}, which helps understand their relative importance.  As shown in the leftmost panel, the central DM and stellar masses of halo G1 peak at
$t_{\rm u} \sim 10.3~ \mathrm{Gyrs}$. This is followed by a sudden of 15\% and 5\% decrease in DM and stellar mass respectively, and a bar forms immediately after that. The bar is associated with the start of the secular DM and stellar mass loss from the central regions that lasts until the present day. Overall, this constitutes a total loss of 29\% (21\%) of the peak DM (stellar) mass, with the AGN responsible for 50\% (21\%) of this decrease and the stellar bar for the remainder. Halo G3 also experiences a disruptive AGN-driven gas blowout, but contrary to halo G1, its bar forms before the blowout occurs. The only mechanism responsible for the DM loss in the hydrodynamical counterpart of halo G2 is the presence of the stellar bar. Similarly to the previous examples (G1 and G3), the formation of the bar in halo G2 at $t_{\rm u} \sim 4 \ \mathrm{Gyrs}$ is associated with the onset of the secular DM and stellar mass loss. Finally, halo G4 never forms a bar nor experiences a major gas blowout, and its inner DM mass remains roughly constant until redshift z=0.

\subsection{AGN-driven gas blowouts}

As we have shown in the previous section, strong gas blowouts caused by AGN activity are able
to induce a substantial decrease in the central stellar and DM
masses. Haloes G1 and G3 experience such blowouts, as evidenced by 
the sharp decrease in their enclosed stellar and dark matter masses
shown in the top panels of Fig.
\ref{figure_2}. These occur at different times, with
the blowout in halo G1 taking place at
$t_{\rm u} \sim 10.3 \ \mathrm{Gyrs}$ and that in halo G3 at
$t_{\rm u} \sim 8 \ \mathrm{Gyrs}$. 

In halo G1, we can see a steady decrease of $\sim 50\%$ of the gas mass
in the central regions before the proper blowout that removes the remaining gas occurs (see vertical dashed line). This prior decrease is
associated with star formation and stellar feedback, which, as we have
verified visually, does not disrupt the gas disc. In fact, despite
this decrease in gas mass, the gas disc becomes more compact and reaches a
surface density comparable to that of the stellar disc just before it
is disrupted. This `compactification' could be caused by the torques
exerted by the galaxy that flew by earlier \citep{Blumenthal.2018}.
Finally, a significant amount of gas mass is fed into the central
black hole, triggering an outburst of the AGN that destroys the
gas disc and removes virtually all gas from the central regions. A
similar process occurs in halo G3, although this galaxy is already
strongly barred prior to the AGN-driven blowout event. The bar could provide
another mechanism to facilitate the inflow of gas towards the centre
of the galaxy \citep[e.g.][]{Sanders.1976,Fanali.2015}.

 \begin{figure}
    \centering
    \includegraphics{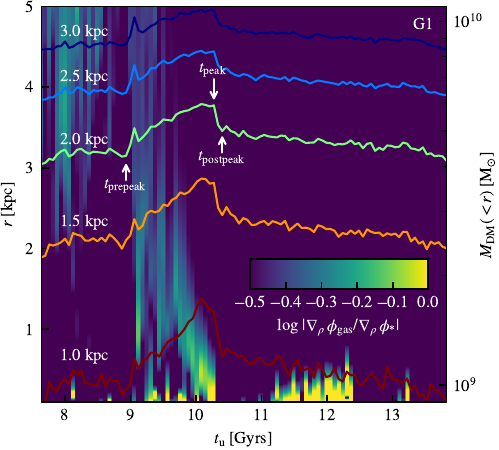}
    \caption{The colour scale indicates the ratio between the gravitational accelerations exerted
      by the gas relative to the stars,
      $\nabla_{\rho} \phi_{\rm gas}/\nabla_{\rho} \phi_{\rm *}$, along
      a random azimuthal direction contained within the midplane of
      the galaxy of halo G1. This is shown as a
      function of time (x-axis) and distance to the centre of the halo
      (y-axis). The solid lines show the evolution of the enclosed dark matter mass within spherical apertures ranging from 1\,kpc to 3\,kpc, in increments of 0.5\,kpc, as indicated by the labels on
      the left hand side.The vertical white arrows indicate
      the visually determined times used to quantify the contraction
      and expansion of both the DM halo and the stellar component
      caused by changes in the gas disc.}
    \label{figure_3}
\end{figure}

A requirement for blowouts to be effective in altering the
distribution of dark matter is that gas must be strongly gravitationally coupled to the dark matter prior to the blowouts \citep{Benitez-Llambay.2019}. We have checked whether this is the case for
halo G1 by comparing the gravitational force exerted by the gas and
the stars along a random azimuthal direction contained within the
midplane of the galaxy. The leftmost top panel of Fig~\ref{figure_2} shows that baryons dominate the central regions of halo G1. Therefore, to asses the importance of gas for the gravitational force, we only need to determine the relative contribution between the gas and stellar components. We compute the force by taking the
gradient of the gravitational potential calculated by direct summation
over all particles within 30~kpc from the centre of the halo. The
choice of azimuthal direction has little effect on the estimated
forces during the time of interest, given that there is no prominent
non-axisymmetric feature prior to the blowout.

Fig.~\ref{figure_3} shows the temporal
evolution of the gravitational force exerted by the gas, relative to that
of the stars, as a function of the distance to the centre. Focusing on the times prior
to the blowout
($9 \, \mathrm{Gyrs} \leq t_{\rm u} \leq 10.3 \, \mathrm{Gyrs}$), the
gas makes a contribution similar to the stars in the innermost regions. This is not the case
for most of the evolution at late times since there is very little gas
left after the blowout. Initially, the gravitational contribution of
the gas disc is significant (but not dominant) throughout the central
5~kpc of the halo. The gas becomes increasingly gravitationally
important in the central regions over time. As mentioned before, this
is due to the fact that the gas disc becomes more compact during this
time. Just before the blowout occurs, the density of the gas disc and the enclosed dark matter and stellar masses peak, which indicates that the baryonic blowout is responsible for the accompanying mass loss in all the components. 

\begin{figure*}
    \centering
    \includegraphics{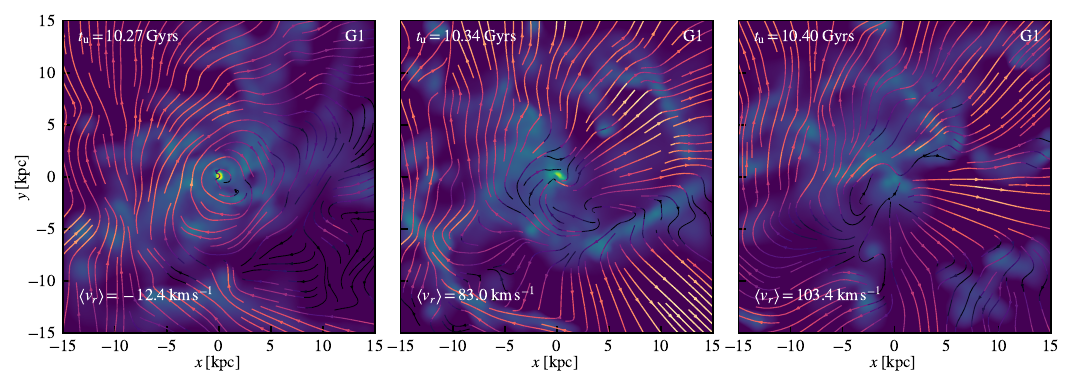}
    \caption{Surface density (colour scale) of the face-on gas content of halo G1, with each panel from left to right showing its distribution before, during and after the AGN outburst, respectively. Streamlines indicate the gas velocity with a colour intensity that is proportional to its magnitude. The average radial velocity of gas within a sphere of 30kpc from the centre of halo G1 is shown in the lower left corner. The times at which the galaxy was imaged are indicated in the top left corners of each panel.}
    \label{figure_4}
\end{figure*}

We have explicitly checked that the large-scale winds are driven by AGN. Three images of the gas content of the galaxy before, during and after the outburst are shown in Fig. \ref{figure_4}. Prior to the start of the event, most gas is concentrated in the centre of the halo, where the black hole resides. It has a net negative average radial velocity, corresponding to inflow that is manifest in the compactification of the gas component observed in Fig.~\ref{figure_3}. Once the outburst commences, most gas is quickly evacuated from the centre in just $\sim$120~Myrs, with outflow velocities that are in excess of 100 $\mathrm{km}\,\mathrm{s}^{-1}$ during the later stages.

Although the gas disc is gravitationally dominant only in the very central
regions prior to being blown out, the sudden gas blowout has a measurable
effect even at radii several times
larger, well beyond where the gravitational contribution of the gas is significant.  The solid lines in Fig.~\ref{figure_3} show
the evolution in time of the enclosed DM mass within apertures ranging
from 1 to 3~kpc. Although the decrease in DM mass is larger at 
smaller radii, where the gravitationally coupling between the dark matter and the gas was larger, it is still detectable at larger radii. This also underlines
the importance of the timescale over which gravitational perturbations
act. The long timescale that it takes for the perturbation to grow ensures that the removal of the gas causes a maximal effect at larger distances from the centre.

To understand why this is the case, consider a system in which all
particles are on circular orbits. When a central perturbation is
exerted, the orbits of particles become elliptical and are able to
come closer to the centre (contraction). When the perturbation is
removed, particles that have descended down the perturbed potential
well will have gained energy, allowing them to reach radii further
beyond their original radius (expansion). Evidently, the
picture is more nuanced in a more realistic scenario in which
particles are not necessarily on circular orbits nor in the same
orbital phase. Nonetheless, this example helps illustrate the
consequences of such perturbations.

Our interpretation implies that the degree of initial contraction and subsequent expansion
must be related: a shell that responds strongly to the
addition of a perturbation will initially contract and subsequently
expand more than one with a weaker response. Thus, the amplitude of
the expansion will, in part, be determined by how strong the initial
contraction was. Secondly, the timescale of the perturbation
determines its effectiveness in altering the kinematics of a radial
shell: if the duration of the perturbation is short compared to the
dynamical timescale of a given radial shell, particles do not have
enough time to deviate significantly from their original orbits and
change their energy. On the other hand, if the timescale is sufficiently
long, its effect becomes maximal.

To illustrate this we investigate whether there is a correlation
between the expansion and contraction of different radial shells of
halo G1, as well as the dependence of their amplitude with distance
from the centre of the halo. We identify the times at which the
gravitational perturbation sourced by the gas disc starts
($t_{\rm prepeak}$) and ends ($t_{\rm postpeak}$). We can thus measure
the enclosed masses at these times relative to the time when they
peaked ($t_{\rm peak}$), and thus estimate the degree of contraction
and expansion of each shell. Although identifying $t_{\rm peak}$ and
$t_{\rm postpeak}$ is straightforward, locating the time at which the
perturbation starts is less so. By visually inspecting the evolution
of the enclosed masses of DM (Fig.~\ref{figure_3}), we can
estimate the time at which the halo starts contracting
($t_{\rm u} \sim 9\mathrm{Gyrs}$). This coincides with the time when
gas was being delivered to the central regions. 

\begin{figure}
    \centering
    \includegraphics{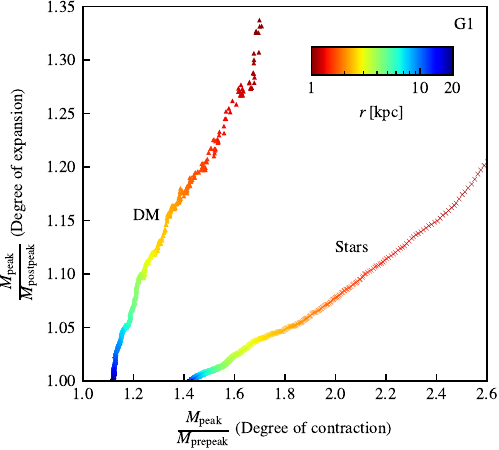}
    \caption{Degree of expansion (vertical axis) as a function of degree of contraction (horizontal axis), for different radial shells in the radius range, $1.0 \lesssim r / \rm kpc \lesssim 20.0$, for halo G1. We estimate the quantities by identifying the time at which the dark matter peaks and measure the mass prior to the peak, $M_{\rm prepeak}$ ($t_{\rm u} \sim 9 \rm \ Gyr$), and after the peak, $M_{\rm postpeak}$ ($t_{\rm u} \sim 10.3 \rm \ Gyr$). These times are indicated in Fig.~\ref{figure_3}. The colour of each marker indicates the value of the spherical aperture.}
    \label{figure_5}
\end{figure}

Once these times have been measured, the degree of
expansion and contraction of each shell is estimated by taking the
ratio of enclosed masses at different apertures, ranging from 1 to
20~kpc. This is shown in
Fig.~\ref{figure_5}, which demonstrates that the degree of expansion is indeed related to the degree of contraction, as expected from our previous arguments. Furthermore, the amplitude increases towards the centre of
the halo, with the DM component losing $\sim 25\%$ and $\sim 10\%$ within 1 and
3~kpc respectively. Similarly, the expansion of the stellar component correlates well with the degree of contraction, although in a different manner compared to the DM component.  Also worth noting is the
offset from unity along the x-axis, likely caused by the growth of both the halo and the galaxy. This is a consequence of our definition of the degree of contraction, which includes the \textit{total} mass increase. Their relative offset is explained by the fact that the galaxy grows more than the DM halo during the timescale under consideration. Finally, differences
between the dynamical properties of the DM and stars might affect the degree of contraction caused by the gas, since circular orbits respond more strongly.

To ascertain how common AGN-fueled blowouts are, we inspect the 
mass evolution of the gas, DM and stars within 2~kpc
from the centre of all MW-mass haloes. We then identify the times when the galaxy loses a large amount of gas in a short timescale (less than 100 Myrs, which corresponds to the time resolution of our data outputs) and inspect whether there is an associated decrease in the stellar
and DM masses. To rule out a localised blowout that is not sufficiently strong to
disrupt the gas disc, the gas content within 5~kpc should have also significantly decreased. Seven haloes out of the 45 studied here show definite evidence supporting that they
experience such events at least once during their lifetime, with two others showing 
strong hints that they did. This comprises roughly 15\% to 20\% of the sample studied 
here, with the rest not experiencing such an event or having a complicated evolution 
that prevents us from making a definite statement on whether they have 
experienced one or not.

These findings are applicable to the real universe as long as strong AGN outbursts are able to completely remove the central gas content of galaxies. The parameters modulating their efficiency in EAGLE were calibrated by requiring that the model should reproduce  a number of population statistics and scaling relations. Nonetheless, this is not the only prescription that meets these requirements. An example is the IllustrisTNG model, which includes a kinetic feedback mode at low mass accretion rates instead of the purely thermal implementation used in EAGLE. These modelling differences lead to different predictions for poorly constrained relations, such as how gas-rich the circumgalactic medium is as a function of halo mass below $\sim 10^{12}\,\Msun$ \citep{Kelly.2021, Davies.2020}. This suggests the EAGLE AGN model is more effective at removing baryons on the MW mass scale than other similarly realistic simulations. It would be interesting to examine whether AGN outbursts similar to the ones discussed above are also present in other simulations using different ways to model AGN feedback.

\subsection{Stellar bars}

In a number of Milky Way mass halos we observe a decrease in the
central mass of dark matter and stars taking place over several Gyrs, as opposed to the AGN-induced gas blowouts, which correspond to Myrs timescales. This decrease occurs in haloes whose central regions are dominated by stars, with the largest decreases associated to strongly barred galaxies. The top panels
of Fig.~\ref{figure_2} show three such examples. Halos
G1 and G2 have a global maximum in the enclosed DM mass followed by a
monotonic decrease such that by $z = 0$, they have lost $\sim 20\%$
and $\sim 40\%$ of the peak DM mass within 2~kpc respectively. This
number excludes the initial, blowout-induced mass decrease observed in
halo G1.  By contrast, halo G3 shows a non-monotonic evolution of the
central mass of DM and stars. This reflects the nature of its past
evolutionary history, which is much more turbulent than for haloes G1
and G2. In contrast to these, which remained relatively undisturbed
once the peak mass had been reached, halo G3 underwent several mergers
and flybys by surrounding galaxies.

The development of a stellar bar in our simulations is
associated with the outward transfer of stellar and DM mass. The
evolution of bars is driven by the exchange of angular momentum with
the surrounding components of the system \citep{Athanassoula.2003}. To
measure the strength of the bar, we first orient the galaxy so that it
is viewed face-on. This is achieved by aligning the spin of the
stellar component, computed by measuring the total angular momentum of
all stars within 5~kpc from the centre, along the $z$-axis. The galaxy
is then split into several concentric cylindrical annuli of 4~kpc in
height and of variable width, such that each encloses 500 stellar
particles. This choice provides better spatial resolution than bins of
constant width in the barred regions, which contain more stellar
particles than the outer regions. For each bin we measure the strength
of the quadrupole moment of the azimuthal distribution of stellar
particles, relative to their monopole strength:
\begin{equation}
    A_{2} \equiv \dfrac{\sqrt{a^{2}_{2} + b^{2}_{2}}}{a_{0}} \, , 
\end{equation}
where:
\begin{align}
    a_{m} &= \sum^{N}_{i}M_{i} \, \mathrm{cos}\, m\phi_{i} \, ,\\
    b_{m} &= \sum^{N}_{i}M_{i} \, \mathrm{sin}\, m\phi_{i} \, .
\end{align}
The sums are taken over all stellar particles in the bin, with $M_{i}$ 
and $\phi_{i}$ their masses and azimuthal angles, respectively. Additionally, we also measure the quadrupole moment phase angle via $\phi_{2} = 0.5 \, \mathrm{arctan}(b_{2}/a_{2})$.

\begin{figure*}
    \centering
    \includegraphics{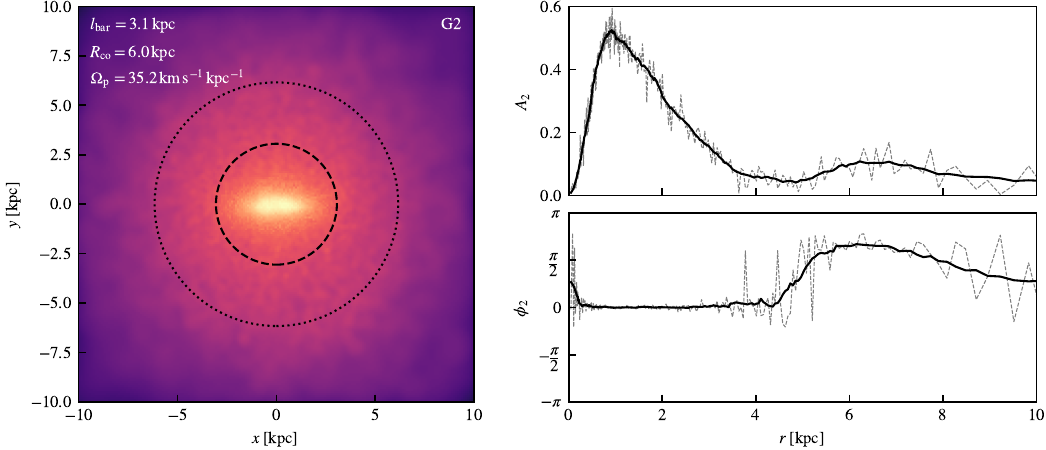}
    \caption{\textit{Left panel}: logarithmic
      surface density (colour scale) of the face-on view of the stellar
      component of halo G2. The face-on orientation is chosen such that the bar is aligned
      along the $x$-axis. The values of the surface density range
      from $10^{10} \, \mathrm{M}_{\odot} \, \mathrm{kpc}^{-2}$ to
      $10^{5} \, \mathrm{M}_{\odot} \, \mathrm{kpc}^{-2}$. The values
      of the bar length, pattern speed and corotation radius are shown in the top left corner. Its extent and corotation
      radius are also indicated by the dashed and dotted circles, respectively. \textit{Top right panel}:
      radial variation of the $A_{2}(r)$ metric for halo G2 at
      $z = 0$. We use the peak value to characterize the strength of
      the bar. The dashed lines show the raw values obtained using
      cylindrical annuli 4~kpc in height, each enclosing 500 stellar
      particles. These values are smoothed using a Savitzky-Golay
      filter (black solid line). \textit{Bottom right panel}: Same as the panel above, but for the quadrupole moment phase angle.}
    \label{figure_6}
\end{figure*}

Finally, a Savitzky-Golay filter is used to smooth the radial
variation of $A_{2}(r)$, with the length of the smoothing window set
to 5\% of the total number of
annuli. The right panels of Fig.~\ref{figure_6} show
the $A_{2}(r)$ profile and the quadrupole moment phase angle for halo G2 at $z = 0$. The former has a prominent
peak within 4~kpc associated with the presence of a strong
bar, which shows a consistent orientation out to 4~kpc. We use the peak value, $A^{\rm max}_{2}$, to
estimate the strength of the bar for each galaxy as a function of time. In this work we use a threshold of $A^{\rm max}_{2} = 0.15$ to estimate when a bar forms, with the choice behind this value being strictly operational. Given that features and artifacts not related to bars can also boost the quadrupole moment in complicated cosmological simulations such as this one, we visually inspect the stellar distribution of galaxies with $A^{\rm max}_{2} \geq 0.15$ to confirm the presence of a bar.

Two other important properties of the bar are its length and pattern
speed. To measure the length, we employ the definition adopted by
\citet{Algorry.2017}: the radius at which $A_{2}(r)$ first drops below
0.15 after it has reached $A^{\rm max}_{2}$. The dashed black circle
in the left panel of
Fig.~\ref{figure_6} shows the extent
of the bar determined in this way.

We measure the bar pattern speed directly by computing the change in
the orientation of the bar between consecutive temporal outputs,
i.e. $\Omega_{p} = \Delta \theta_{p}/\Delta t$. The angle of the bar, $\theta_{p}$,
is measured from the phase of the quadrupole moment at the radius
where $A_{2}(r)$ peaks.

The bottom panels of Fig.~\ref{figure_2} show the
time evolution of the strength of the stellar bars in the galaxies
illustrated on the top panels. The bars in both halo G1 and G2 form at
around the time when the central DM mass begins to decrease. Once
formed, the bars generally increase in strength monotonically. 
The evolutionary story of halo G3
is less trivial and the bar goes through periods of strengthening and
weakening. At early times ($t_{\rm u} \leq 7 - 7.5\,\mathrm{Gyrs}$), the value of $A_{2}$ is large but in spite of this, we do not visually recognise a bar. The large value of $A_{2}$ is caused by mergers occurring along the line-of-sight, and by the fact that the centre becomes ill-defined during this period. The end result is that the projected stellar distribution exhibits a large quadrupole moment. Additionally, transient elongations of the stellar distribution influence the values of $A^{\rm max}_{2}$, as seen in the variations of its value in the bottom panel of Fig.~\ref{figure_2}.
The newly formed bar weakened significantly 
probably as a result of the rapid increase in the density of the central gas
that preceded the AGN blowout. Interestingly, most of this gas
was distributed along a bar-like feature aligned with the stellar
bar. After increasing in strength again, there is an additional weakening episode that is likely caused by a merger event \citep{Ghosh.2021}. In all three cases, the bars have significantly slowed down
by $z = 0$, with the ratio of corotation radius to bar
length equal to 2.1, 1.9 and 2.0 respectively.

For halos G1 and G2 the secular decrease in the central DM mass is
clearly associated with the formation and subsequent evolution of the
bar. Nonetheless, even though both have strong bars of similar
strengths at $z = 0$, each lost different amounts of mass from the
central regions. This is due to the fact that the bars formed at
different times and evolved by different amounts. To investigate
further the connection between the decrease in central DM mass and the
prominence of the stellar bar, the age of the bar should be taken into
account. The metric we use is the average mass loss rate between
$z_{\rm peak}$ and $z = 0$, normalised by the peak DM mass within 2~kpc.  We consider only those examples that exhibit a monotonic
decrease in the central DM mass, e.g. haloes G1 and G2, but not G3. This selection was done by visually inspecting the central mass evolution of all the MW mass haloes in the simulation.
Twenty-three out of the initial 45 haloes satisfy this criterion,
which preferentially selects galaxies that have undergone relatively
undisturbed evolution after the peak in central DM mass was
reached. Finally, we exclude sudden DM mass decreases associated with
the gas blowouts discussed in the previous section, since here we are
interested in the bar-driven secular decrease.

\begin{figure}
    \centering
    \includegraphics{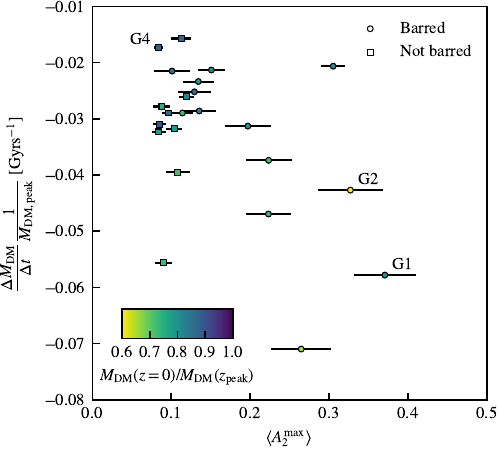}
    \caption{The time-averaged fractional mass loss rate of haloes that 
      exhibit a monotonically decreasing central DM mass as a
      function of their average bar strength. The average bar strength
      is computed from $z=0$ until the time when the enclosed dark matter peaks. The circles show galaxies
      that have values of $A^{\rm max}_{2} \geq 0.15$ for longer than
      1~Gyrs. These are assigned to the sample of barred galaxies. Galaxies that do not satisfy this criterion are
      assigned to the unbarred population. The horizontal error bars
      show the spread in values of $A^{\rm max}_{2}(t)$ for each
      galaxy since $t_{\rm peak}$, and indicate how much the
      quadrupole moment strength has evolved since then. Finally, each dot is
      coloured according to present-day to peak mass ratio, as indicated in the legend.} 
    \label{figure_7}
\end{figure}

Fig.~\ref{figure_7} shows the variation of
the average fractional mass loss rate between $z_{\rm peak}$ and $z=0$
of each monotonically expanding halo as a function of their
time-averaged bar strength,
$\langle A^{\rm max}_{2} \rangle$\footnote{The reason behind using the
  average bar strength instead of the $z = 0$ strength is that some of
  the galaxies had (weak) bars in the past that later
  dissolved.}. This value was computed by averaging $A^{\rm max}_{2}$
from the time at which the enclosed DM peaked up to $z = 0$. The
horizontal error bars indicate the variation of $A^{\rm max}_{2}(t)$,
with low values corresponding to galaxies whose quadrupole
moment strength remained relatively unchanged (e.g. if the stellar
disc was axisymmetric throughout the simulation or the bar did not
weaken or strengthen). We further classify galaxies into barred or
unbarred depending on whether they had a value of $A^{\rm max}_{2}$
greater than 0.15 during at least 1~Gyrs. This
ensures that even galaxies that were barred in the past but are not at
$z = 0$ are correctly identified, whilst excluding high, transient
values of $A^{\rm max}_{2}$.

Broadly speaking, Fig.~\ref{figure_7}
suggests that the stronger the time-averaged bar strength of a galaxy
is, the greater its secular DM mass loss rate. To 
quantify this, we calculate Pearson's correlation
coefficient $\mathcal{R}$ between $\langle A^{\rm max}_{2} \rangle$
and the (average) fractional mass loss rate for both populations. 
The median value for the barred sample is $\mathcal{R} = -0.7^{+0.2}_{-0.2}$, giving support to our previous claim. On the other hand, the unbarred sample has a median of $\mathcal{R} = 0.2^{+0.3}_{-0.3}$, which is consistent with no correlation. It is worth noting that the strongest correlation is found between how much the bar evolved over time and the time-averaged fractional mass loss rate. The correlation coefficient between these two variables is $\mathcal{R} = -0.83\pm 0.07$. The quoted uncertainties were obtained using a bootstrap technique.

Fig.~\ref{figure_7} shows several other
interesting features. Firstly, a number of haloes with very low values
of $\langle A^{\rm max}_{2} \rangle$ have a wide range of mass loss
rates. These galaxies were never barred, and thus the expansion could
not have been caused by a bar. We find that these galaxies were gas rich in the past, which caused the haloes to contract. As time progressed, star formation locked some of baryons in stars but the resulting supernovae feedback expelled gas from the central regions. As a result, the overall baryon mass decreased over time, leading to a slightly less contracted halo at later times. This reduction in the central baryon content takes place on much longer timescales than the AGN phase.
Secondly, the moderately barred galaxy above the G2 data point has a very low
central mass loss rate given its bar strength. As indicated by the
horizontal error bars, there was very little change in
$A^{\rm max}_{2} $ which remained roughly constant at
$A^{\rm max}_{2} \sim 0.3$ since its formation. This hints at the need
for evolution in the strength of a bar for effective transfer of
central DM mass outwards.  Finally, to reiterate the importance of the
timescale over which the stellar bar acts, halo G1 experienced a
greater mass loss rate than halo G2 but lost less central DM
mass. This is because G1 had a bar for $\sim 4 \ \mathrm{Gyrs}$,
whereas G2 had it for $\sim 10 \ \mathrm{Gyrs}$.

Finally, many of the barred galaxies considered here also exhibit a decrease
in the central mass of stars, although to a lesser extent than their
DM content. This is observed in all three barred examples shown in the
top panels of Fig.~\ref{figure_2}. The presence of a
bar is expected to not only affect the dark matter, but also the
stellar distribution. Whether particles near resonances are able to
emit or absorb angular momentum is dependent on their dynamical
properties. Spheroid components such as a dark matter halo or a
stellar bulge are net absorbers of angular momentum, so one might expect the stellar
bulge to be similarly affected. We have checked whether this is the case for halo G2 by classifying its $z=2$ stars into bulge and disk components based on their circularities, $\epsilon_{\rm circ} \equiv j_{\rm z}/j_{\rm z,circ}(E)$. By tracking a subset of the most bound particles of each component, we found that the stellar mass loss within 2~kpc is dominated by expansion of the stellar bulge. The net effect is in line with
the `smoothing' effect that non-axisymmetric features have on the
rotation and mass distribution curves of disc galaxies
\citep{Berrier.2015}. Indeed, the rotation curves of these galaxies
are strongly peaked at small radii when the bar initially formed but
are less so as time progresses.

These findings are in qualitative agreement with those of \citet{Algorry.2017}, although there are differences between these studies. Firstly, we analyse a higher resolution version of the EAGLE simulation, with an increase of almost one order of magnitude in the particle mass resolution. This allows us to study the evolution of the inner regions of haloes more confidently.
On the other hand, the smaller volume reduces the number of barred galaxies that we can study. Nonetheless, we are able to have a more detailed look at the evolution of the central regions of these haloes. This reveals the importance of the age of the bar, which is primarily determined by the assembly history of its halo. Even for galaxies  with similar  bar strength at $z=0$, substantial differences in formation time alter how much the halo de-contracts.  Moreover, we also note that the expansion of the dark matter halo is often accompanied by an expansion of the stellar component. Lastly, \citet{Algorry.2017} used a stellar mass criterion to select their sample, whereas we select ours  based on the virial mass of the dark matter halo. In practice, this results in the stellar component of the haloes in our study being less massive than those in  \citet{Algorry.2017}. This follows from the fact that in EAGLE the stellar mass function is underestimated \citep{Schaye.2015}, so one has to consider more massive haloes to find sufficiently massive galaxies.

\subsection{Halo contraction}

All the haloes considered in this study end up having a higher central
density compared to their DMO counterparts. This is a consequence of the
accumulation of baryonic mass resulting from the dissipational
collapse of gas during the assembly of the galaxy, the so-called
`adiabatic contraction' \citep{Barnes.1984,Blumenthal.1986,Cautun.2020}. This trend is opposed by
the processes we have discussed here which reduce the central DM
density. We now consider how the bar-driven secular evolution of a galaxy alters
the degree of contraction of its host halo, bearing in mind that the
degree of contraction depends on the halo assembly history
\citep{Abadi.2010}.

An illustrative example is halo G2, which had the strongest and
longest-lived bar in our sample. This galaxy formed $50\%$ of its
stars by $z = 2.6$ and was left largely undisturbed for a large
fraction of the age of the universe. Changes to its density profile
are solely driven by internal, secular processes such as the influence
of its stellar bar. To estimate how contracted the halo is relative to
its DM counterpart, we follow the procedure described by
\citet{Abadi.2010}. Firstly, we define a shell of radius, $R_{\rm i}$,
in the DMO simulation that encloses a given amount of dark matter,
rescaling the particle masses:
$M^{\rm DMO}_{\rm DM}(<R_{\rm i}) = (1-f_{\rm b})M^{\rm DMO}_{\rm
  tot}(<R_{\rm i})$. We then find the corresponding radius,
$R_{\rm f}$, in the hydrodynamical simulation that encloses the same
amount of dark matter, i.e.
$M^{\rm hydro}_{\rm DM}(<R_{\rm f}) = M^{\rm DMO}_{\rm DM}(<R_{\rm i})
$. In practice, this amounts to enclosing equal numbers of DM
particles for each shell. Once these two radii have been found, we
measure the total enclosed masses,
$M_{\rm tot} = M_{\rm DM} + M_{\rm b}$, for each, where $M_{\rm b}$ is
the mass of baryons. The ratio $R_{\rm f} / R_{\rm i}$ measures the
degree of contraction ($R_{\rm f} / R_{\rm i} < 1$) or expansion
($R_{\rm f} / R_{\rm i} > 1$) of a radial shell as a function of the
increase ($M^{\rm i}_{\rm tot} / M^{\rm f}_{\rm tot} < 1$) or decrease
($M^{\rm i}_{\rm tot} / M^{\rm f}_{\rm tot} > 1$) of total mass
contained within it. This allows a straightforward comparison to the
predictions of the simple adiabatic contraction model, for which
$R_{\rm i}M^{\rm i}_{\rm tot} = R_{\rm f}M^{\rm f}_{\rm tot}$.

\begin{figure}
    \centering
    \includegraphics{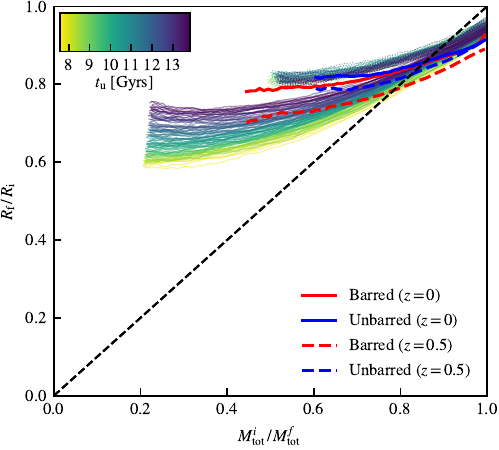}
    \caption{Response of the dark matter halo to the assembly of the
      galaxy at its centre. The ratio $R_f/R_i$ is shown as a function of the
      change in enclosed total mass $M^{\rm i}_{\rm tot}/M^{\rm f}_{\rm tot}$ at several
      times for halo G2 (solid coloured lines) and a halo hosting an unbarred 
      galaxy (similar to G4; dotted coloured lines). The values are only shown for radial 
      shells with radii larger than the convergence radius defined by \citep{Power.2003}, which is different for 
      each halo. For a consistent  comparison, we show only values measured at $t_{\rm u} \geq 7.5\mathrm{Gyrs}$. The solid and dashed lines show the
      $z = 0$ and $z = 0.5$ average values for the barred (red) and unbarred (blue) sample of haloes shown in Fig.~\ref{figure_7}, respectively. The average only includes haloes that are relaxed. We do not show the average value 
      for $M^{\rm i}_{\rm tot}/M^{\rm f}_{\rm tot} \leq 0.45$  $(0.60)$  because only a small 
      number of haloes hosting barred (unbarred) galaxies reach these low values. The diagonal dashed line shows the 
      prediction of the adiabatic contraction model.}
    \label{figure_8}
\end{figure}

Fig.~\ref{figure_8} shows how these ratios
vary as a function of time for halo G2, as well as for an unbarred
galaxy with a very small reduction in its central DM content. The
values are only shown for shells with radii larger than the
\cite{Power.2003} convergence radius of the DMO halo, with each shell
containing 100 more DM particles than the previous one. For
consistency when comparing the evolution between the chosen haloes,
only the last $\sim 6\,\mathrm{Gyrs}$ of the simulation outputs are shown. Consequently, the evolution of
halo G2 is larger than shown in here, since its expansion began at
$t_{\rm u} \sim 4 \, \mathrm{Gyrs}$. Nonetheless, it is evident that
its contraction evolves strongly with time. This is in contrast to the
unbarred halo, which remains virtually unchanged over the 
plotted time period.

As discussed in other studies \citep[e.g.][]{Abadi.2010, Cautun.2020},
the adiabatic contraction model overestimates the degree of
contraction in the central regions. We find that it also underpredicts
it at larger radii as a result of baryonic outflows. Consider a radial
shell at the virial radius of the halo, $R_{200}$. At such large
distances, contraction should be negligible and thus
$R_{\rm f} / R_{\rm i} \sim 1$.  Under the assumption of no DM shell
mixing, the values of $M^{\rm DMO}_{\rm tot}(<R_{\rm i})$ and
$M^{\rm hydro}_{\rm tot}(<R_{\rm f})$ only depend on the enclosed baryonic
mass. For the collisionless case, this is simply
$M^{\rm DMO}_{\rm b}(<R_{\rm i}) = f_{\rm b}M^{\rm DMO}_{\rm tot}(<R_{\rm i})$. However,
if feedback expels baryons beyond the virial radius, then the baryon
fraction at such a distance is less than the cosmic baryon
fraction. In other words,
$M^{\rm DMO}_{\rm b}(<R_{\rm i}) \geq M^{\rm hydro}_{\rm b}(<R_{\rm f})$ and thus
$M^{\rm DMO}_{\rm tot}(<R_{\rm i}) / M^{\rm hydro}_{\rm tot}(< R_{\rm f}) \geq
1$. Evidently, this is violated if there is a reduction in the
enclosed DM mass at a fixed physical aperture caused by the reduced
growth of the halo associated with the loss of baryons at early times,
as discussed by \cite{Sawala.2013}. However, this is not the case for
MW-mass halos in the EAGLE simulations
\citep{Schaller.2015}. 

Focusing on the sample of monotonically expanding haloes,
Fig.~\ref{figure_8} also shows how the
average values of the two ratios plotted change between $z = 0.5$
(red) and $z = 0$ (black). When computing the averages, only relaxed
halos are included. To identify relaxed halos we adopt the most
restrictive condition proposed by \cite{Neto.2007}: the centre of mass
should be offset from the centre of potential by less than
$0.07R_{\rm vir}$. We only show the averages for
$M^{\rm i}_{\rm tot}/ M^{\rm f}_{\rm tot} \geq 0.45 $ $(0.60)$ for the barred (unbarred) sample, since very few
haloes reach smaller values. The barred sample is, on average, more contracted than the unbarred one as a consequence of their more massive stellar components. The average contraction of each population has decreased over time; the evolution of the unbarred sample is less  than that of the barred one.

The systematic shift in the average contraction of the haloes in our
sample is due to the secular processes discussed earlier. Since one of
the driving effects is stellar bars, which transfer angular momentum
from the stars to the dark matter particles (see Appendix A), this process could lead
to different dark matter particle distribution functions in halos
whose central galaxies have a bar compared to those which do
not. These differences may be relevant for distribution function-based
models of halo contraction (e.g. \citealt{Callingham.2020}).

Finally, there is considerable halo-to-halo scatter introduced by
several factors such as the mass of the central galaxy, the assembly
history, the orbital distribution of DM particles, etc. Among these,
AGN-driven blowouts could also play a role due to their stochastic
nature. Given the small size of our halo sample, we can only note this
trend which needs to be confirmed by larger simulations.

\section{Discussion and conclusions}

In this work we have studied the evolution of the central distribution
of dark matter in simulated Milk Way mass halos, drawn from a
$\Lambda$CDM cosmological hydrodynamics simulation. Specifically, we
have investigated how the mass within the inner 2 - 5~kpc is affected by processes
associated with the growth of the central galaxy in the halo.  We analysed
45 haloes taken from the high resolution version of the EAGLE
simulation, selected by requiring that their final mass, $M_{200}$, be
similar  to that of the Milky Way at $z = 0$.

As in previous studies, we find that, at the present day, the halos
are more centrally concentrated than their counterparts in a DMO
simulation. However, the
degree of contraction is significantly less than expected in the
simple adiabatic
contraction model \citep{Blumenthal.1986} and its refinements
\citep{Gnedin.2004,Cautun.2020}. We also find that there are times during the
evolution of a halo when its central dark matter mass decreases,
although it always remains more massive than its DMO twin.  We have
identified two main processes responsible for lowering the central
dark matter mass.

The first is AGN-induced gas blowouts. These
events involve gas that had slowly become so dense so as to become gravitationally dominant in the central regions. As the gas is violently expelled, the
central regions of the halos expand in a process analogous to that
discussed by \cite{Navarro.1996a}. Both dark matter and stars
participate in the expansion. Interestingly, we find that these blowouts can reduce the enclosed dark matter and stellar mass at radii much larger than those in which the gas is gravitationally dominant. Although it is clear that the effect of the blowouts fades away at larger distance, we hypothesize that the very long timescales of the gravitational perturbations caused by the gas before being blown out are responsible of this effect. In all cases, the initial inflow of gas occurs during or shortly after an interaction with a nearby galaxy.

The second process that causes the central halo regions to expand is a
bar-mediated transfer of angular momentum. This transfer reduces the
central DM density at a rate that is likely set by the time-averaged stellar
bar strength. The net change in the central dark matter density
depends on the length of time since the bar formed. 

The effects of AGN-driven gas blowouts and bar-driven angular momentum
transfer that we have investigated in this work are not confined to
the central regions of the halos and can be seen out to at least 5~kpc
from the centre. Here we have focused attention on the inner 2~kpc
since this radius lies in the well-converged region as judged by the
\cite{Power.2003} criterion
($r_{\rm power} \sim 1.5\mathrm{kpc} - 1.75\mathrm{kpc}$).

Not all the galaxies in our sample undergo the two processes just
described. As Fig. \ref{figure_2} shows, there is significant halo-to-halo scatter due to a variety
of factors including differences in the assembly history of the halo,
the central mass of the galaxy and likely the orbital distribution of dark matter particles. Roughly 30\% of the studied sample host bars. About 15\% to 20\% of our MW-mass haloes have experienced at least one AGN
blowout capable of reducing their central stellar and dark matter
densities. This is likely to be a lower limit, as we focus on haloes
with well defined evolutionary histories. At high redshifts, when the haloes and galaxies are assembled, it is difficult to assess the importance of the AGN blowouts.

While the reduction in central dark matter mass due to the presence
of stellar bars similar to those that form in EAGLE is likely to be
generic, the reduction caused by AGN-driven blowouts is expected to be specific to
the EAGLE subgrid model. It would be interesting to explore if similar
effects are present in other hydrodynamics simulations.

The processes discussed in our paper indicate that the assembly of baryons in Milky Way-size haloes induces a complicated reaction in the DM halo. The degree of DM contraction in these haloes cannot be solely characterised by the present-day baryonic distribution, but by their complicated past evolutionary history. Our own Milky Way contains two of the ingredients that source the complexity highlighted in our study: a stellar bar and a supermassive black hole at the centre. Our results suggest that unless the baryonic effects described in our paper are taken into account, studies that rely on contraction-based models, such as mass estimates of the Milky Way, or direct or indirect searches of dark matter, could contain biases that are very difficult to account for.

\section*{Acknowledgements}
ABL acknowledges support by the European Research Council (ERC) under the European Union's Horizon 2020 research and innovation programme (GA 757535) and UNIMIB's Fondo di Ateneo Quota Competitiva (project 2020-CONT-0139). CSF, SMC and VJFM  acknowledge support by the European Research Council (ERC)
through Advanced Investigator grant DMIDAS (GA 786910) and Consolidated Grant ST/T000244/1. This work used the DiRAC@Durham facility managed by the Institute for Computational Cosmology on behalf of the STFC DiRAC HPC Facility (www.dirac.ac.uk). The equipment was funded by BEIS capital funding via STFC capital grants ST/K00042X/1, ST/P002293/1, ST/R002371/1 and ST/S002502/1, Durham University and STFC operations grant ST/R000832/1. DiRAC is part of the National e-Infrastructure.
\section*{Data Availability}

Instructions on how to access the EAGLE database that contains the data used in this paper can be found at  \url{http://icc.dur.ac.uk/data/ }.


\bibliographystyle{mnras}
\bibliography{references}

\appendix

\section{Angular momentum evolution}

Stellar bars mediate the transfer of angular momentum between different components of the system, with the net flow dependent on the morphology of the latter. We investigate this in the context of this work by tracking a subset of the most bound stellar and DM particles belonging to halo G2. This criterion selects particles which occupy the central regions of the system, where the effect of the bar will be the strongest.


We classify stellar particles as bulge or disk-like using the method of \citet{Abadi.2003}, which is based on their circularity parameters:
\begin{equation*}
    \epsilon_{\rm circ} \equiv \dfrac{J_{\rm z}}{J_{\rm z,circ}(E)} \, .
\end{equation*}
$J_{\rm z}$ is the angular momentum component parallel to the stellar disk spin and $J_{\rm z,circ}(E)$ is that of a circular orbit with the same binding energy. Thus, $\epsilon_{\rm circ}$ ranges from +1 for co-rotating circular orbits to -1 for counter-rotating ones. $J_{\rm z,circ}(E)$ is obtained by computing the binding energy and $J_{\rm z}$ of all stars in the system. Only certain regions in this phase space are accessible, with its bounds corresponding to $J_{\rm z,circ}(E)$. Here we assign particles with $\epsilon_{\rm circ} \geq 0.9$ to the disk, and those with $\epsilon_{\rm circ} \leq 0.5$ to the bulge. 

\begin{figure}
    \centering
    \includegraphics{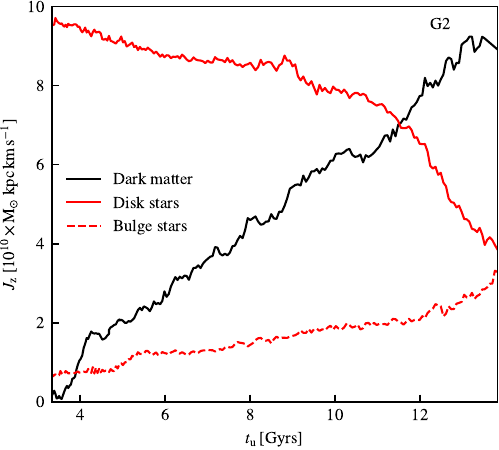}
    \caption{Angular momentum evolution of the component parallel to the stellar disk spin for the ($z = 2$) 5000 most bound particles of each of the components of halo G2. This is shown for disk and bulge stars by the red solid and red dashed lines, respectively. The black line corresponds to the dark matter component, smoothed using a linear Savitzky-Golay filter.}
    \label{appendix_figure_1}
\end{figure}
Finally, we identify the ($z = 2$) 5000 most bound particles of each of the components present in halo G2. This is just before its stellar bar formed. We then track the same particles over time and follow the evolution of their angular momentum component parallel to that of the stellar disk spin. This is shown in Fig.~\ref{appendix_figure_1}. We see that the disk subset steadily loses angular momentum, whereas the DM and bulge gain it, as expected. Given that we track a subset of the whole system, this is not a closed system and thus angular momentum is not strictly conserved. There are likely other sources of angular momentum not accounted for in this analysis, such as newly formed stars and the gas disk. Nonetheless, this gives a qualitative view on how angular momentum is redistributed by the bar.


\bsp	
\label{lastpage}
\end{document}